**Origins and dissociation of pyramidal <c + a> dislocations in magnesium and its alloys**

Zhigang Ding[a], Wei Liu*[a], Hao Sun[a], Shuang Li[a], Dalong Zhang[b],

Yonghao Zhao[a], Enrique J. Lavernia[b], Yuntian Zhu[a,c]

[a]Nano and Heterogeneous Materials Center, School of Materials Science and Engineering, Nanjing University of Science and Technology, Nanjing, Jiangsu 210094, China

[b]Engineering and Materials Science, University of California, Irvine, CA 92697-1000, USA

[c]Department of Materials Science and Engineering, North Carolina State University, Raleigh, NC 27695, USA

∗Corresponding author; Email: weiliu@njust.edu.cn

**Abstract**

Alloying magnesium (Mg) with rare earth elements such as yttrium (Y) has been reported to activate the pyramidal <c + a> slip systems and improve the plasticity of Mg at room temperature. However, the origins of such dislocations and their dissociation mechanisms remain poorly understood. Here, we systematically investigate these mechanisms using dispersion-inclusive density functional theory, in combination with molecular dynamics simulations. We find that <c + a> dislocations form more readily on the pyramidal I plane than on the pyramidal II plane in Mg. The addition of Y atoms in Mg facilitates the dissociation of <c + a> dislocations on pyramidal II, leading to the easier formation of the pyramidal II than pyramidal I in Mg-Y alloy. Importantly, in pyramidal II slip plane, a flat potential-energy surface (PES) exists around the position of stable stacking fault energy (SFE), which allows cooperative movement of atoms within the slip plane. Alloying Mg with Y atoms increases the range of the PES, and ultimately promotes different sliding pathways in the Mg-Y alloy. These findings are consistent with experimentally observed activation of the pyramidal II <c + a> slip system in Mg-Y alloys, and provide important insight into the relationship between dislocation structure and macroscopic enhancement of plasticity.

# 1. Introduction

Magnesium (Mg) and its alloys are promising candidate materials for energy efficient transportation vehicles and devices due to their high strength and light weight [1-4]. However, the plasticity of Mg at room temperature is poor, due to their basal-type texture and limited number of slip systems available to accommodate applied plastic deformation [5-7]. Activation of <c + a> dislocations on pyramidal plane is an effective way to improve the plasticity of Mg, because these dislocations can accommodate *c*-axis strain and provide sufficient independent slip systems. As such, many experimental [8-13] and theoretical studies [6, 14, 15] have been devoted to achieving a better understanding on the formation mechanisms and dissociation modes of <c + a> dislocations in Mg and its alloys.

The <c + a> dislocations can slip on either pyramidal I (P1) {10-11} or pyramidal II (P2) {11-22} planes [16], along with different atomic densities, activation barriers, and dislocation cores. Notably, no consensus has been achieved on the prevalent slip planes in Mg and its alloys. For example, from the TEM tilting experiments the P2 <c + a> dislocation in single-crystalline Mg was suggested as the dominant slip plane [17-19]. Using the same methodology, slips on both P1 and P2 plane were detected in compressed Mg micropillar [20-22], but the prevalent slip plane was not identified in these studies. Compressing single-crystal Mg along its *c*-axis revealed that the P1 <c + a> dislocation is the dominant non-basal slip mode in Mg, since as the critical resolved shear stress (CRSS) of <c + a> slip on P1 is lower than that on P2 [9]. The debate on the prevalent slip plane also exists in theoretical studies. Molecular dynamics (MD) simulations showed that <c + a> dislocations predominately nucleate on P1 slip plane from free surface and cavities, during *c*-axis compression and tension simulations [14, 23]. However, P2 slip plane was found to be the preferred slip plane by discrete dislocation dynamics simulations [24, 25]. Although experiments have clearly demonstrated that <c + a> slip can be activated in Mg under *c*-axis compression [9, 23], the CRSS for slips on pyramidal plane is still significantly larger than that on basal plane at room temperature, which is difficult to activate in pure Mg.

Recent experiments have demonstrated that alloying rare earth elements can significantly improve the plasticity of Mg at room temperature by activating pyramidal <c + a> dislocations [5, 6, 26-29]. By comparing Mg and Mg-Y alloys, Sandlöbers *et al.* [5] observed large amount of basal stacking faults (both $I_1$ and $I_2$) and pyramidal <c + a> dislocations in Mg-Y alloys. They thus argued that $I_1$ stacking fault acts as the heterogeneous nucleation source of <c + a> dislocation, and alloying Y in Mg facilitates the nucleation of the <c + a> dislocation. Zhang *et al.* [28] also suggested that enhanced <c + a> slip benefits the plasticity of Mg-Y alloys, but the creation of $I_1$ faults originates from a side effect from the <c + a> slip. A very recent study explored the possible relationship between $I_1$ stacking faults and <c + a> dislocations, but did not elaborate the non-conservative formation mechanisms of copious $I_1$ stacking faults [30]. Also, recent MD simulated results showed that Y atoms increase the CRSS of basal slip more than that of non-basal slip, eventually reducing the difference in the CRSS between different slip systems [31].

The <c + a> dislocations on both P1 and P2 plane may dissociate into partials at the position of stable stacking fault energy (SFE) [10-13]. Early experiments predicted that at room temperature a P2 <c + a> full dislocation would dissociate into two partials on the P2 plane [17]. Recent experiments provided further evidence that the <c + a> dislocation dissociates into two equal partials on the P2 plane [32]. The similar dissociation modes observed from experiments have been supported by density-functional theory (DFT) and MD simulations, along with a large discrepancy in the position of stable SFE (from 0.35b to 0.5b) [12, 13, 15, 33]. MD simulations also revealed other dissociation modes, e.g., dissociate into partial <c> and <a> dislocations [11, 34], or into two partials along the <20-23> direction [30]. Notably, recent MD studies have reported that the <c + a> dislocation can dissociate into three basal-dissociated immobile dislocation structures [2]. These immobile dislocations would hinder the motion of other *c*-axis associated dislocations, and results in high hardening and low plasticity of magnesium [4]. Accordingly, for further development of high plasticity Mg alloys, it is critical to understand the origin and behavior of such dislocations to improve the plasticity of Mg alloys at room temperature.

In this paper, we use DFT calculations, in combination with MD simulations, to systematically study the GSFE curves of <c + a> slip on both P1 and P2 plane for Mg and Mg-Y alloys. We find that <c + a> dislocation can form more readily on the pyramidal I plane than on the pyramidal II plane in Mg. The addition of Y atoms in Mg facilitates the dissociation and formation of <c + a> dislocations on P2, via decreasing both the unstable SFE and stable SFE of pure Mg. Specifically, for the P2 slip plane, when a dislocation dissociates at stable position, atoms on the slip plane are not on the lattice position and a flat potential-energy surface (PES) exists around this position, which allows atoms to move cooperatively towards their lattice position. Our MD simulations further revealed that the movement of atoms induces stacking fault cooperative movement (SFCM) from off-lattice positions to their lattice positions. Based on such SFCM, we provide a symmetric sliding pathway for P2 <c + a> dislocations, in which the trailing partial dislocation starts from this lattice position like the inverse process of the leading partial dislocation. Our calculations also find that alloying Y increases the moving range of SFCM and reduces the SFE in P2 <c + a> dislocation. These fundamental findings provide physical understanding of the formation mechanisms and slip features of P2 dislocations in HCP metals, and may help with resolving the discrepancy between experiments and simulations: experiments found superimposed Y atoms can promote the P2 <c + a> slip, whilst DFT predicted the impeding influence. Our results help clarify the relationship between observed dislocation structures and macroscopic enhanced room temperature plasticity by alloying Mg with Y.

## 2. Computation approach

All DFT calculations performed here were completed using the Vienna Ab-initio Simulation Package (VASP) code [35]. The interaction between the valence electrons and ionic cores was described by the projector augmented wave (PAW) method [36]. The standard Perdew-Burke-Ernzerhof (PBE) form of the generalized gradient approximation (GGA) [37] was used as the exchange-correlation functional throughout the paper. The optB88-vdW exchange-correction functional [38-41] was utilized to account for dispersion interactions in our calculations. In recent work [42], we carried out optB88-vdW calculations to systematically study the slip mechanisms of pure Mg slabs, and demonstrated the promising role of van der Waals forces in the slipping process and the plasticity of Mg.

Using the lattice constant of a = 3.198 Å and c = 5.194 Å [42], we established a 12-layer slab with 48 atoms for pyramidal slip plane (see Fig. 1). Each slab was separated by 10 Å vacuum to eliminate artificial interactions between slabs. One Mg atom in the sixth layer was substituted by an alloying element to model the Mg alloy systems. For slab calculations, the Brillouin zone was sampled with a 6 × 10 × 2 $k$-point mesh, along with an energy cutoff of 400 eV. A residual force threshold of 0.001 eV/Å was used for geometry optimizations. The convergence tests showed that the error bar of the total energy was less than 0.1 meV/atom when more accurate settings were used in our computations. As to the GSFE curve calculations, we used the conventional direct crystal slip technique, in which some coordinates of the atoms are constrained during structure optimizations. Three structure optimization methods were taken into account in our computations: 1) relax only z coordinates of each atom (termed as "z-relax"); 2) relax both x and z coordinates ("xz-relax"); and 3) fully relax all coordinates ("full-relax").

The MD simulations were performed using the large-scale atomic/molecular massively parallel simulator (LAMMPS) code [43]. Interatomic interactions between Mg atoms were described by an MEAM-type potential [44]. The lattice constants at 300 K tested by this potential are a = 3.210 Å and c = 5.210 Å, in excellent agreement with experimental results. Note that these values are larger than those from the optB88-vdW functional, which were calculated at 0 K. We established cell with [-1100] × [11-23] × [11-20] direction that contains 8600 atoms for MD simulations. The Nose-Hoover thermostats were used to maintain constant temperature. Isothermal-isobaric ensemble (NPT) was employed for an independent relaxation along three directions; the temperature of the system was kept at 300 K. The structures of perfect crystal and stable stacking fault position (0.5b) were relaxed for 10000 steps within a total time scale of 10 ns.

## 3. Results and discussion

### 3.1 GSFE curves of Mg

We first determine the GSFE curves of Mg along <11-23> direction slip on P1 and P2 planes. Following the z-relax method for GSFE calculations, we shift the upper half of the crystal with respect to the lower half of the crystal along the <11-23> direction. Figure 2(a) shows the calculated GSFE curves for both P1 and P2 slip plane, from which both stable SFE and global unstable SFE on P1 plane are lower than those of P2. Our finding is consistent with experimental observation that the formation of the <c + a> slip is easier on P1 than on the P2 plane [9]. The position of stable SFE is 0.45b in P1 and 0.35b in P2, which corresponds to the position of dislocation dissociation. However, this dissociation modes of P2 is distinct with experimental [32, 45] results, that P2 <c + a> slip will dissociate into two equal partials.

The above calculations can reproduce reasonably well the value and position of stable SFE in Refs. [12, 13, 15, 33]. However, when applying this z-relax optimization approach to corrugated faults, such as stacking faults on pyramidal planes, less accurate or even incorrect position for stable SFE would be obtained [12, 46, 47]. Notably, allowing atoms to relax in slip plane has been reported and verified in recent studies of the stable SFE on some slip planes of HCP metals [48-55]. Unlike the z-relax method, the additional relaxation in the x-direction allows extra accommodation motions and coordinate glide that are perpendicular to the slip direction [32]. We re-calculated the GSFE curves by using the xz-relax method. As shown in Fig. 2b, atoms relaxed along x direction reduce the global unstable SFE value from 365 to 315 mJ/m$^2$ in P1 plane, and from 411 to 327 mJ/m$^2$ in P2. Moreover, our computed results showed that the position of stable SFE becomes 0.4b and 0.5b for P1 and P2, respectively. Also, the magnitude of the stable SFE reduces from 203 to 163 mJ/m$^2$ in P1, and from 213 to 158 mJ/m$^2$ in P2. These results indicate that under xz-relax condition, local atom movement permitted by the additional relaxation leads to minimum values and change the position of stable SFE. These results agree with the conclusions from the experimental and other theoretical results, *i.e.*, the <c + a> dislocation is easier to form on P1 than on P2 [9, 56-58]. Importantly, this stable position on P2 plane is consistent with the experimental results that pyramidal II <c + a> dislocation will dissociate into two equal partials [32].

When comparing the GSFE curves calculated by PBE with those by optB88-vdW functionals, we also find that van der Waals forces have significant influence on the motion of dislocations. Our prior calculations of <a> slip on different slip planes revealed that the van der Waals interactions have significant influence on the unstable SFE, although their influence on stable SFE is less [42]. However, analysis of the GSFE curves of <c + a> demonstrates that the van der Waals interactions have obvious influence on both unstable SFE and stable SFE, although the feature of the related curves from PBE and optB88-vdW are similar. This implies that the van der Waals interactions critically affect both the slip process as well as the dissociation of pyramidal dislocations.

### 3.2 GSFE curves of Mg-Y alloy

To analyze the influence of Y on the formation and dissociation of <c + a> dislocations, we now study the GSFE curves of Mg-Y alloy on both P1 and P2 planes using the z-relax and xz-relax methods. In the calculations, one Mg atom in the sixth layer was substituted by Y to model the Mg-Y alloy systems. We computed the GSFE curves of Mg and Mg-Y alloys that are obtained with the z-relax method on two slip systems. Consistent with previous studies [12, 13, 15, 33], the GSFE curves of Mg-Y alloy are initially lower, then higher than that of Mg once beyond the position of local unstable SFE (Fig. 3a). Remarkably, the stable SFE of Mg-Y disappeared in both planes, which suggests that the dissociation of <c + a> dislocation would be impeded by alloying with Y. Nevertheless, alloying Y maximizes the global unstable SFE from 365 to 389 mJ/m$^2$ in P1 and 411 to 496 mJ/m$^2$ in P2, which prevent the formation of <c + a> dislocations in both planes.

We re-calculated the GSFE curves of Mg-Y alloy by the xz-relax method. As shown in Fig. 3b, the global unstable SFE values of P1 and P2 are almost identical (346 *vs.* 356 mJ/m$^2$), suggesting that <c + a> dislocations are allowed to exist simultaneously in both planes. Note that our theoretical calculations agree nicely with the experimental findings of Sandlöbes *et al.* [5], showing that both P1 and P2 <c + a> dislocation can be observed in Mg-3 wt.% Y alloy. These curves also illustrate that the stable SFE of Mg-Y vanishes in P1, which prevents the formation of <c + a> dislocations. In contrast, the stable SFE remains in P2 (141 mJ/m$^2$ at 0.45b with xz-relax method), indicating that the <c + a> dislocation on P2 is more readily to dissociate than that on the P1 plane in Mg-Y alloys. Although the stable SFE and local unstable SFE have lower magnitude, the increasing global unstable SFE (from 327 to 346 mJ/m$^2$) would prevent the formation of <c + a> dislocation in Mg-Y alloy. More importantly, however, this process still conflicts with the experiment findings, *i.e.*, extensive <c + a> slip has been documented to occur in Mg-Y alloy [5, 27].

### 3.3 Position of stable stacking fault on pyramidal II slip plane

The accurate position of stable SFE on pyramidal planes is particularly important for understanding the core structure, formation, and dissociation of <c + a> dislocations. In DFT calculations, optimization of structures with full atomic relaxation allows us to identify the position of stable SFE on slip plane [58]. Here, we focus on the position of stable SFE on P2 plane, due to the fact that the <c + a> dislocation on P2 is easier to dissociation than on P1. We fully relaxed the structure of Mg and Mg-Y alloy on P2 plane at 0.35b, 0.45b and 0.5b, respectively, and found that positions of stable SFE from the full-relax method are the same as those from the xz-relax method.

To understand the difference in the position of stable SFE, we now analyze the charge density distributions of Mg and Mg-Y at 0.5b on P2 slip plane by using different optimization approaches. As shown in Fig. 4a, similar charge density distributions can be seen for Mg, no matter whether the system was relaxed with xz-relax or full-relax. In this case, charge densities aggregate more around the slip plane as compared to those from the z-relax method. This suggests that the optimization approach of xz-relax and full-relax can give lower energy, which is closely related to the position of stable SFE. Moreover, Fig. 4b clearly illustrates that the position of stable SFE changes from 0.5b to 0.45b upon alloying.

More importantly, the optimized structures at stable position show that atoms move cooperatively in the slip plane (Figs. 4c-4d). This phenomenon is similar to the "atomic shuffling motion" proposed in Ref. [33]. In that work, large-scale ab-initio calculations of <c + a> pyramidal screw dislocations in Mg prove that when atoms on a pyramidal plane are shifted to their stable positions, they end up at off-lattice positions, under the influence of "atomic shuffling motion", reaching their lattice position after relaxation [33]. However, they did not notice the cooperative movement phenomenon, which will reduce the SFEs and change the sliding pathway. In our case, the above atomic motion phenomenon suggests that there exists a flat PES around the position of stable SFE, which allows atoms to move cooperatively on the slip plane. This atomic motion may further change the slip modes of pyramidal <c + a> slip systems.

To further demonstrate the cooperative movement and the existence of the flat PES, we carried out classic MD simulations. The structures of perfect crystal and stable stacking fault position (0.5b) were relaxed for 10000 steps within a total time scale of 10 ns. Figure 5a shows the moving trajectory of perfect crystal at different relaxation time, no obvious atom move can be seen from the trajectory. In the relaxation progress, atoms basically vibrate slightly at their original positions, which are presumably because of the thermal vibrations around the center of the atoms; even relaxed with long time, no change will occur in the trajectory. In contrast, in the moving trajectory of the stable stacking fault position structure (Fig. 5b), atoms move cooperatively around their position. Compare the structures of 0 fs with those of 70 fs, we can see that atoms on slip plane move away from their original positions. After relaxing 140 fs, atoms on the next-nearest slip plane also started to move. As time goes on, atoms on layers that are far away from slip plane even started to move. Importantly, the atoms on the same slip plane move cooperatively around their original position, which suggests a flat PES at the stable SFE. Atoms move on this flat PES induces stacking fault migrate from off-lattice position to their lattice position on the slip plane.

### 3.4 Dissociation mode of pyramidal II <c + a> dislocation

We now discuss the reaction of P2 <c + a> dislocation. The existence of stable position suggests that a dislocation will dissociate into two partials at this location. The 1/3<11-23> dislocation can dissociate into two partials according to the following reaction,

$$\frac{1}{3}<11\text{-}23> \rightarrow \lambda \frac{1}{6}<11\text{-}23> + (1-\lambda)\frac{1}{6}<11\text{-}23> \tag{1}$$

where λ is a constant coefficient depends on the position of the stable SFE. Different λ values were proposed in literature [12, 13, 48]. The λ values of our calculation are 0.5 for P2 slip plane. However, the asymmetry and larger global unstable SFE value in Mg-Y implies that further slip along this direction will be impeded, which will prevent the P2 <c + a> slip. Note that the two partial dislocations are along the same direction, which was recently evidenced by dislocation core structures obtained using both DFT and MD simulations [44]. This dissociation is different from that of FCC and basal dissociation in HCP metals, in which the dissociation are along different directions.

More importantly, based on the SFCM on the stable SFE, the pyramidal <c + a> dislocation can form by following three steps: first, leading partial dislocation slip along 1/3<11-23> direction arrives at the position of stable SFE; second, stacking fault migrates cooperatively from off-lattice position to the lattice position; third, trailing partial dislocation starts from the lattice position, just like the inverse process of first process, and does not need to go through the large global unstable SFE (Fig. 3b). The GSFE curves and the slip processes are shown in Fig. 6. In this context, the 1/3<11-23> dislocation dissociates into two partial dislocations by the following reaction,

$$\frac{1}{3}<11\text{-}23> \rightarrow \lambda \frac{1}{6}<11\text{-}23> + \text{SFCM} + (1-\lambda)\frac{1}{6}<11\text{-}23> \qquad (2)$$

In this reaction, a P2 <c + a> dislocation dissociates into two symmetrical partials. Importantly, this process is energetically favorable and conservative in terms of the slip process. This leads to a significant reduction in the global unstable SFE of the trailing partial dislocation (see the second peak in Fig. 2). We find that the superimposed Y atoms decrease both the unstable SFE and stable SFE of pure Mg, which facilitates the formation of pyramidal II <c + a> dislocation. Particularly, alloying with Y can increase the range of SFCM, which directly improves the plasticity of Mg.

When applying this mechanism to Mg alloys with other elements (Al, Ca, Tb, Er, Dy, and Ho), we also find similar stable SFE position and SFCM phenomenon. The positions of stable SFEs are around the middle point, and the range of flat PES is differs with that of Mg-Y alloys. As shown in Fig. 6, the width of the PES of Mg is increased by alloying with Ca, Tb, Er, Dy, and Ho, but is reduced by alloying with Al. These results are in agreement with the experiment observation that in Mg-(Ca, Tb, Er, Dy, and Ho) alloys there are large number of <c + a> dislocations [26], but few in Mg-Al alloy [31]. On the basis of our analysis, we conclude that both the range of PES and unstable SFE have significant influence on the formation of pyramidal II <c + a> dislocations.

## 4. Conclusions

We have carried out density-functional theory calculations to systematically study the slip and dissociation of pyramidal <c + a> dislocations in Mg and its alloys. Our calculations find that <c + a> dislocations is easier to form on the pyramidal I (P1) plane than on the pyramidal II (P2) plane in Mg, while the addition of Y atoms in Mg facilitates the dissociation of <c + a> dislocations on P2. Our calculations also find that at the position of stable SFE there exists a flat potential-energy surface (PES) that induces stacking fault cooperative movement (SFCM). When dislocation dissociates at stable position, atoms at stable position can move back to their lattice positions from the off-lattice positions by the influence of SFCM. Correspondingly, further dislocation slip starting from the lattice positions may slip like the inverse process of first <c + a>/2 dislocation. Alloying with Y element can increase the range of the PES and decrease both unstable and stable SFE, which reasonably explain the appearance of large number of <c + a> dislocations and enhancement of the room temperature plasticity of Mg−Y alloy. This dissociation mechanism may be applicable to many other Mg alloys, such as Mg-Al, Mg-Ca, Mg-Tb, Mg-Er, Mg-Dy, and Mg-Ho. Our results provide new insight into the relationship between observed dislocation structure and enhanced room temperature plasticity by alloying with Y and helpful in the design high plasticity Mg alloys.

**Acknowledgements**

This work was supported by the National Key R&D Program of China (Grant 2017YFA0204403). W.L. is grateful for support from the NSFC (51722102, 21773120), and the Fundamental Research Funds for the Central Universities (30917011201). Y.T.Z. acknowledges the support of the Jiangsu Key Laboratory of Advanced Micro-Nano Materials and Technology and the US Army Research Office (W911 NF-17-1-0350). D.Z. and E.J.L. acknowledge support from NSF-CMMI-1437327.

**References**

[1] T. M. Pollock, Weight loss with magnesium alloys, Science 328 (2010) 986-987.

[2] Z. Wu, W. Curtin, The origins of high hardening and low ductility in magnesium, Nature 526 (2015) 62-67.

[3] Z. Wu, W. A. Curtin, Mechanism and energetics of <c + a> dislocation cross-slip in hcp metals, Proc. Natl. Acad. Sci. 113 (2016) 11137-11142.

[4] Z. Wu, B. Yin, W. A. Curtin, Energetics of dislocation transformations in hcp metals, Acta Mater. 119 (2016) 203-217.

[5] S. Sandlöbes, S. Zaefferer, I. Schestakow, S. Yi, R. Gonzalez-Martinez, On the role of non-basal deformation mechanisms for the ductility of Mg and Mg-Y alloys, Acta Mater. 59 (2011) 429-439.

[6] S. Sandlöbes, M. Friák, J. Neugebauer, D. Raabe, Basal and non-basal dislocation slip in Mg-Y, Mater. Sci. Eng. A 576 (2013) 61-68.

[7] M. Yoo, Slip, twinning, and fracture in hexagonal close-packed metals, Metall. Trans. A 12 (1981) 409-418.

[8] S. Agnew, J. Horton, M. Yoo, Transmission electron microscopy investigation of <c + a> dislocations in Mg and α-solid solution Mg-Li alloys, Metall. Mater. Trans. A 33 (2002) 851-858.

[9] K. Y. Xie, Z. Alam, A. Caffee, K. J. Hemker, Pyramidal I slip in c-axis compressed Mg single crystals, Scr. Mater. 112 (2016) 75-78.

[10] Y. Minonishi, S. Ishioka, M. Koiwa, S. Morozumi, M. Yamaguchi, The core structure of 1/3<11-23> {11-22} edge dislocations in hcp metals, Philos. Mag. A 43 (1981) 1017-1026.

[11] B. Li, E. Ma, Pyramidal slip in magnesium: Dislocations and stacking fault on the {10-11} plane, Philos. Mag. 89 (2009) 1223-1235.

[12] Z. Pei, L. F. Zhu, M. Friák, S. Sandlöbes, J. von Pezold, H. Sheng, C. P. Race, S. Zaefferer, B. Svendsen, D. Raabe, Ab initio and atomistic study of generalized stacking fault energies in Mg and Mg-Y alloys, New J. Phys. 15 (2013) 043020.

[13] T. Nogaret, W. Curtin, J. Yasi, L. Hector, D. Trinkle, Atomistic study of edge and screw <c + a> dislocations in magnesium, Acta Mater. 58 (2010) 4332-4343.

[14] Y. Tang, J. A. El-Awady, Formation and slip of pyramidal dislocations in hexagonal close-packed magnesium single crystals, Acta Mater. 71 (2014) 319-332.

[15] M. Ghazisaeidi, L. Hector, W. Curtin, First-principles core structures of <c + a> edge and screw dislocations in Mg, Scr. Mater. 75 (2014) 42-45.

[16] D. Hull, D. J. Bacon, Introduction to dislocations, Butterworth- Heinemann, 2001.

[17] J. F. Stohr, J. P. Poirier, Etude en Microscopie Electronique du Glissement pyramidal {11-22} <11-23> dans le Magnesium, Philo. Mag. 25 (1972) 1313-1329.

[18] F. Lavrentev, Y. A. Pokhil, Relation of dislocation density in different slip systems to work hardening parameters for magnesium crystals, Mater. Sci. Eng. 18 (1975) 261-270.

[19] T. Obara, H. Yoshinga, S. Morozumi, {11-22} <11-23> slip system in magnesium, Acta Metall. 21 (1973) 845-853.

[20] E. Lilleodden, Microcompression study of mg (0001) single crystal, Scr. Mater. 62 (2010) 532-535.

[21] C. M. Byer, B. Li, B. Cao, K. T. Ramesh, Microcompression of single- crystal magnesium, Scr. Mater. 62 (2010) 536-539.

[22] B. Li, P. Yan, M. Sui, E. Ma, Transmission electron microscopy study of stacking faults and their interaction with pyramidal dislocations in deformed mg, Acta Mater. 58 (2010) 173-179.

[23] H. Fan, J. A. El-Awady, Towards resolving the anonymity of pyramidal slip in magnesium, Mater.

Sci. Eng. A 644 (2015) 318-324.

[24] Z. H. Aitken, H. Fan, J. A. El-Awady, J. R. Greer, The effect of size, orientation and alloying on the deformation of AZ31 nanopillars, J. Mech. Phys. Solids 76 (2015) 208-223.

[25] H. Fan, S. Aubry, A. Arsenlis, J. A. El-Awady, The role of twinning deformation on the hardening response of polycrystalline magnesium from discrete dislocation dynamics simulations, Acta Mater. 92 (2015) 126-139.

[26] S. Sandlöbes, Z. Pei, M. Friák, L. F. Zhu, F. Wang, S. Zaefferer, D. Raabe, J. Neugebauer, Ductility improvement of Mg alloys by solid solution: Ab initio modeling, synthesis and mechanical properties, Acta Mater. 70 (2014) 92-104.

[27] B. Wu, Y. Zhao, X. Du, Y. Zhang, F. Wagner, C. Esling, Ductility enhancement of extruded magnesium via yttrium addition, Mater. Sci. Eng. A 527 (2010) 4334-4340.

[28] D. Zhang, L. Jiang, J. M. Schoenung, S. Mahajan, E. J. Lavernia, TEM study on relationship between stacking faults and non-basal dislocations in Mg, Philos. Mag. 95 (2015) 3823-3844.

[29] D. Zhang, H. Wen, M. A. Kumar, F. Chen, L. Zhang, I. J. Beyerlein, M. Schoenung, S. Mahajan, E. J. Lavernia, Yield symmetry and reduced strength differential in Mg-2.5Y alloy, Acta Mater. 120 (2016) 75-85.

[30] S. Agnew, L. Capolungo, C. Calhoun, Connections between the basal $I_1$ growth fault and <c + a> dislocations, Acta Mater. 82 (2015) 255-265.

[31] K. H. Kim, J. B. Jeon, N. J. Kim, B. J. Lee, Role of yttrium in activation of <c + a> slip in magnesium: An atomistic approach, Scr. Mater. 108 (2015) 104-108.

[32] A. Kumar, B. M. Morrow, R. J. McCabe, I. J. Beyerlein, An atomic-scale modeling and experimental study of <c + a> dislocations in Mg, Mater. Sci. Eng. A 695 (2017) 270-278.

[33] M. Itakura, H. Kaburaki, M. Yamaguchi, T. Tsuru, Novel cross-slip mechanism of pyramidal screw dislocations in magnesium, Phys. Rev. Lett. 116 (2016) 225501.

[34] B. Li, E. Ma, Zonal dislocations mediating {10-11} <10-1-2> twinning in magnesium, Acta Mater. 57 (2009) 1223-1235.

[35] G. Kresse, J. Furthmller, Efficient iterative schemes for ab initio total energy calculations using a plane-wave basis set, Phys. Rev. B 54 (1996) 11169-11186.

[36] P. Blöchl, Projector augmented-wave method, Phys. Rev. B 50 (1994) 17953.

[37] J. P. Perdew, K. Burke, M. Ernzerhof, Generalized gradient approximation made simple, Phys. Rev.

Lett. 78 (1996) 3865.

[38] J. Klimeš, D. R. Bowler, A. Michaelides, Chemical accuracy for the van der Waals density functional, J. Phys.: Condens. Matter 22 (2009) 022201.

[39] J. Klimeš, D. R. Bowler, A. Michaelides, Van der Waals density functionals applied to solids, Phys. Rev. B 83 (2011) 195131.

[40] M. Dion, H. Rydberg, E. Schröder, D. C. Langreth, B. I. Lundqvist, Van der Waals density functional for general geometries, Phys. Rev. Lett. 92 (2004) 246401.

[41] A. Tkatchenko, D. S. R. Jr, R. Car, M. Scheffler, Accurate and efficient method for many-body van der Waals interactions, Phys. Rev. Lett. 108 (2012) 1577-1581.

[42] Z. Ding, W. Liu, S. Li, D. Zhang, Y. Zhao, E. J. Lavernia, Y. Zhu, Contribution of van der Waals forces to the plasticity of magnesium, Acta Mater. 107 (2016) 127-132.

[43] S. Plimpton, Fast parallel algorithms for short-range molecular dynamics, J. Comput. Phys. 117 (1995) 1-19.

[44] Z. Wu, M. Francis, W. Curtin, Magnesium interatomic potential for simulating plasticity and fracture phenomena, Modell. Simul. Mater. Sci. Eng. 23 (2015) 015004.

[45] Q. Yu, L. Qi, R. K. Mishra, J. Li, A. M. Minor, Reducing deformation anisotropy to achieve ultrahigh strength and ductility in mg at the nanoscale, Proc. Natl. Acad. Sci. 110 (33) (2013) 13289-13293.

[46] Y. Minonishi, S. Ishioka, M. Koiwa, S. Morozumi, M. Yamaguchi, The core structures of a <11-23> {11-22} edge dislocation under applied shear stresses in an hcp model crystal, Philos. Mag. A 45 (1982) 835-850.

[47] D. Bacon, M. Liang, Computer simulation of dislocation cores in hcp metals interatomic potentials and stacking-fault stability, Philos. Mag. A 53 (1986) 163-179.

[48] J. Morris, J. Scharff, K. Ho, D. Turner, Y. Ye, M. Yoo, Prediction of a {11-22} hcp stacking fault using a modified generalized stacking-fault calculation, Philos. Mag. A 76 (1997) 1065-1077.

[49] Y. Dou, J. Zhang, Effects of structural relaxation on the generalized stacking fault energies of hexagonal-close-packed system from first-principles calculations, Comp. Mater. Sci. 98 (2015) 405-409.

[50] P. Kwaśniak, P. Śpiewak, H. Garbacz, K. J. Kurzydłowski, Plasticity of hexagonal systems: Split slip modes and inverse peierls relation in α-Ti, Phys. Rev. B 89 (2014) 144105.

[51] P. Kwaśniak, H. Garbacz, K. Kurzydlowski, Solid solution strengthening of hexagonal titanium

alloys: Restoring forces and stacking faults calculated from first principles, Acta Mater. 102 (2016) 304-314.

[52] M. Ghazisaeidi, D. Trinkle, Core structure of a screw dislocation in Ti from density functional theory and classical potentials, Acta Mater. 60 (2012) 1287-1292.

[53] E. Clouet, Screw dislocation in zirconium: An ab initio study, Phys. Rev. B 86 (2012) 144104.

[54] N. Chaari, E. Clouet, D. Rodney, First-principles study of secondary slip in zirconium, Phys. Rev. Lett. 112 (2014) 075504.

[55] N. Chaari, E. Clouet, D. Rodney, First order pyramidal slip of 1/3<1-210> screw dislocations in zirconium, Metall. Mater. Trans. A 45 (2014) 5898-5905.

[56] A. Moitra, S. G. Kim, M. Horstemeyer, Solute effect on basal and prismatic slip systems of Mg, J. Phys.: Condens. Matter 26 (2014) 445004.

[57] J. Zhang, G. Liu, X. Wei, Strengthening and ductilization potentials of nonmetallic solutes in magnesium: First-principles calculation of generalized stacking fault energies, Mater. Lett. 150 (2015) 111-113.

[58] B. Yin, Z. Wu, W. Curtin, Comprehensive first-principles study of stable stacking faults in hcp metals, Acta Mater. 123 (2017) 223-234.

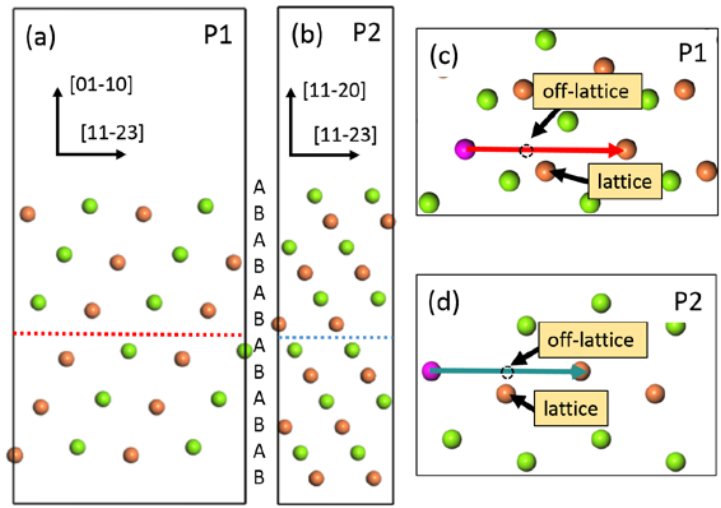

Figure 1: The slab supercells used to calculate the GSFE curves of {1011} <11-23> (P1) slip system (a) and {11-22} <11-23> (P2) slip system (b), the dotted line represents the slip plane. Figure (c) and (d) are the top view structures of both slip systems, and the arrows denote slip direction <11-23>. Orange and green spheres denote A and B packing sequences of the HCP lattice, respectively. To calculate the GSFE curves of Mg-Y alloys, one Mg atom on slip plane was substituted by Y atom (purple sphere). In (c) and (d), the empty dash circle indicates the position of off-lattice position, where the <c + a> dislocation would dissociate, and may move back to lattice position after atomic collective movement.

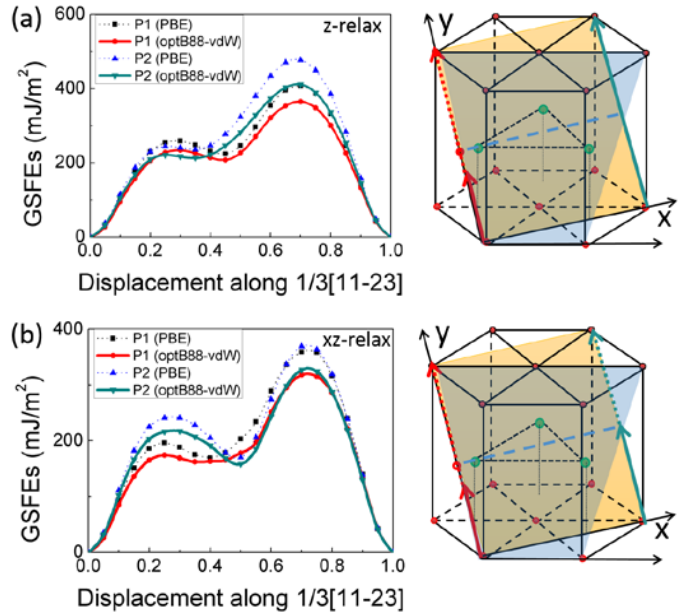

Figure 2: Generalized stacking fault energy (GSFE) curves of {10-11}<11-23> (P1) and {11-22}<11-23> (P2) slip systems of Mg, which were computed from (a) z-relax method and xz-relax method (b). The minimum of each curve corresponds to the stable SFE; the maximum in the curves corresponds to the global unstable SFE. The schematics of sliding pathway are shown on the right of the curves. The red arrows, and blue arrows correspond to slip on pyramidal I and pyramidal II plane; solid arrows and dash arrows correspond to leading and trailing partial dislocation, respectively.

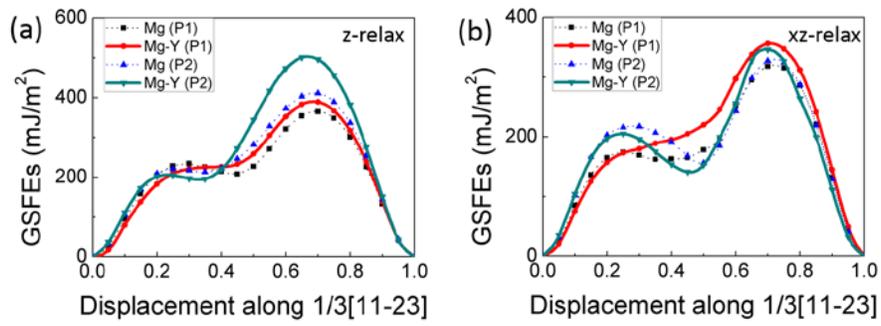

Figure 3: Generalized stacking fault energy (GSFE) curves of {10-11}<11-23> (P1) and {11-22}<11-23> (P2) slip systems of Mg-Y alloy, which were computed from (a) z-relax method and xz-relax method (b). The minimum of each curve corresponds to the stable SFE; the maximum in the curves corresponds to the global unstable SFE.

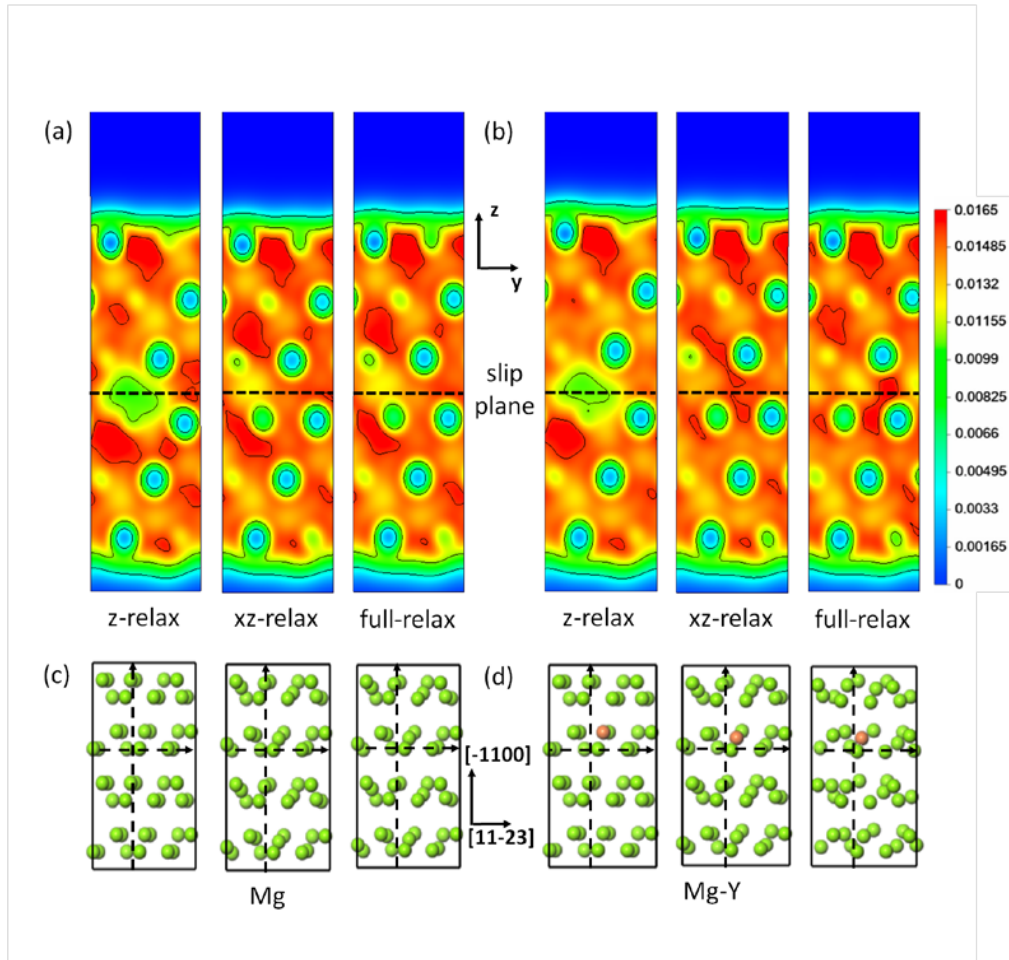

Figure 4: Structures and charge density distributions of (a, c) Mg and (b, d) Mg-Y at 0.5b with three relaxation approaches. In the plots of charge density distributions, red color denotes the area with higher charge density; the dash line represents the slip plane. For both Mg and Mg-Y, the charge density distribution calculated via optimized the structure by full-relax has more charge transfer to the slip plane, which implicates lower energy and the structure corresponds to that of stable position.

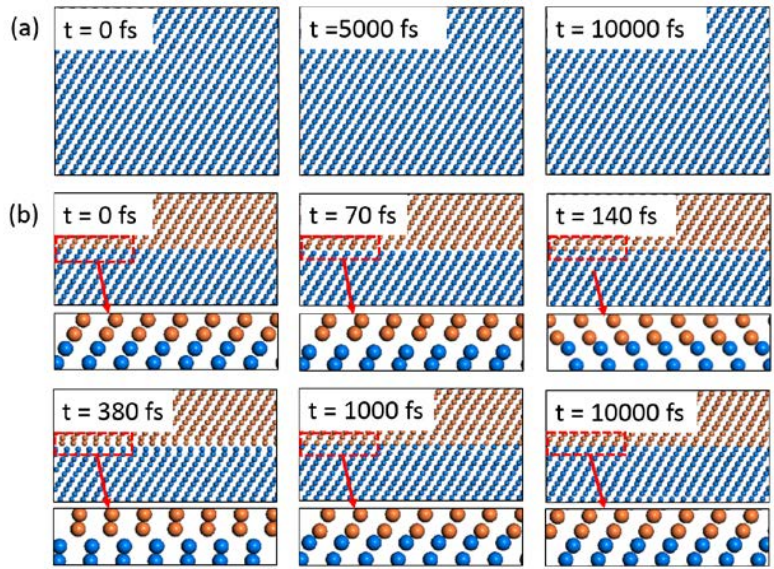

Figure 5: Structures of perfect crystal (a) and stable stacking fault position (b) of Mg at different relaxation time by molecular dynamics simulations. In the perfect crystal atoms vibrate slightly around their original positions and no obvious deformation was detected. In the stable stacking fault structure, atoms on the same slip plane move cooperatively and have uniform move trends around their original position, showing a flat potential energy surface and inducing stacking fault migration from off-lattice position to lattice position. The orange and blue spheres represent atoms up and down of the slip plane.

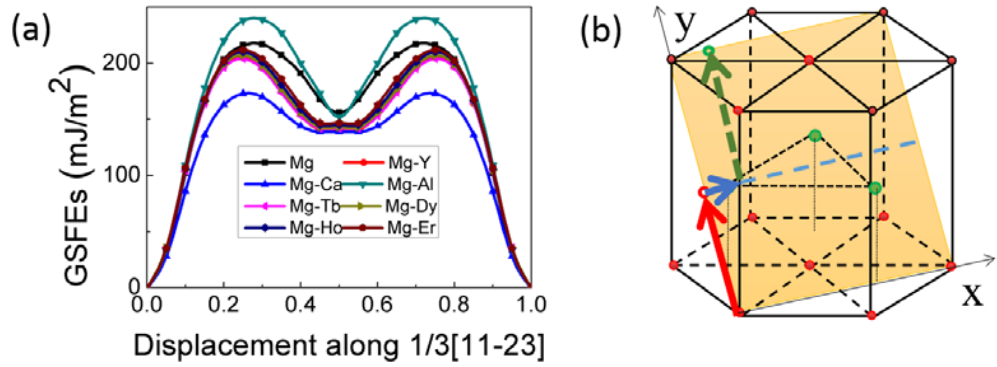

Figure 6: (a) The symmetric GSFE curves for Mg and Mg alloys that involve the stacking fault cooperative movement; the global unstable SFE value is reduced in this case. (b) The schematics of sliding pathway of pyramidal II <c + a> dislocation. The red solid arrow, green dashed arrow, and blue solid arrow correspond to the leading partial dislocation, trailing partial dislocation, and the stacking fault cooperative movement, respectively.